\documentclass[tighten]{aastex63}
\usepackage{amssymb,amsmath,framed}

\usepackage{bm}
\expandafter\ifx\csname package@font\endcsname\relax\else
 \expandafter\expandafter
 \expandafter\usepackage
 \expandafter\expandafter
 \expandafter{\csname package@font\endcsname}
\fi
\usepackage{savesym}
\savesymbol{tablenum}
\usepackage{siunitx}
\restoresymbol{SIX}{tablenum}

\def\bq{\begin{equation}}
\def\eq{\end{equation}}
\def\bqy{\begin{eqnarray}}
\def\eqy{\end{eqnarray}}

\def\p{\partial}

\def\p{\partial}

\def\R{\mathbb{R}}

 \def\q#1#2{q^{\hspace{1pt}#1}_{\,,\hspace{1pt}#2}}
 \def\a#1#2{a^{\hspace{1pt}#1}_{\,,\hspace{1pt}#2}}
 \def\d#1#2{\delta^{\hspace{1pt}#1}_{\,,\hspace{1pt}#2}}

\shorttitle{Minimum Safe Distances for DE-STAR Space Lasers}
\shortauthors{Hibberd}

\begin{document}

\title{\large{Minimum Safe Distances for DE-STAR Space Lasers}}

\correspondingauthor{Adam Hibberd}
\email{adam.hibberd@i4is.org}


\author[0000-0003-1116-576X]{Adam Hibberd}
\affiliation{Initiative for Interstellar Studies (i4is), 27/29 South Lambeth Road London, SW8 1SZ United Kingdom}

\begin{abstract}
The prospect of phased laser arrays in space has received considerable attention in recent years, with applications to both planetary defence and space exploration. The most detailed investigation conducted into such a design is that of the DE-STAR phased array, standing for \textbf{D}irected \textbf{E}nergy \textbf{S}ystems for \textbf{T}argeting of \textbf{A}steroids and explo\textbf{R}ation. DE-STAR is a square modular design which exploits the energy created by banks of solar cells in space to generate and amplify the power of a laser beam. A specific DE-STAR design is expressed as DE-STAR n, where 'n' (typically in the range 0 - 4) equates to the log to base 10 of the side, in metres, of a square bank of lasers. With a DE-STAR 4 structure (10 $\si{km}$ $\times$ 10 $\si{km}$ square) capable of generating a laser beam on the order of tens of gigawatts, clearly there is the potential for such an asset to be deployed as a weapon by targeting locations on Earth. This naturally leads to the question of what effective ways can this possible misuse be removed or at least mitigated, to ensure these powerful space lasers can only be used for their intended purpose, and never malevolent reasons. One solution would be to locate the DE-STAR far enough away so that the laser flux at Earth would be too low. Results indicate that given they should lie 1 $\si{au}$ from the Sun, there are feasible locations for DE-STAR 0-2 arrays where there is no danger to Earth. For DE-STAR 4-5, such is their power, safety measures other than those considered here would have to be adopted. Positions in the Solar System where the DE-STAR lasers have no direct line-of-sight with Earth tend to be unstable, and would require regular corrections using an on-board propulsion system, or preferably using push-back from the laser itself. 
\end{abstract}



\section{Introduction}
\label{sec1}
Lasers are exploited by humanity in outer space for a variety of purposes, amongst these are \citep{BOHACEK2022}:
\begin{itemize}
    \item{\textit{Space traffic management (STM)} where satellite laser ranging (SLR) using ground-based laser infrastructure can for instance monitor space debris and provide alerts for potential collisions, with future planned space-based lasers for satellite tracking.}
    \item{\textit{Space resources utilization (SRU)} an example of which  is Laser-Induced Breakdown Spectroscopy (LIBS), installed on robotic rovers on worlds beyond Earth such as Mars.}
    \item{\textit{Space Exploration and Planetary Defence} the subject of this paper.}
\end{itemize}
In the '60s, at the height of the Space Race, the \textit{United Nations Office for Outer Space Affairs} (UNOOSA) formulated the \textit{Treaty on Principles Governing the Activities of States in the Exploration and Use of Outer Space, including the Moon and Other Celestial Bodies}, known colloquially as the \textit{Outer Space Treaty}. This treaty, spelling out the rules governing the exploitation and exploration of space was accepted and signed in 1967, and now has 115 parties across the globe. Of particular relevance for this paper is article IV which is quoted in full below \citep{Outerspace}: \\

\textit{``States Parties to the Treaty undertake not to place in orbit around the earth
any objects carrying nuclear weapons or any other kinds of weapons of mass
destruction, install such weapons on celestial bodies, or station such weapons in
outer space in any other manner.
 The moon and other celestial bodies shall be used by all States Parties to the
Treaty exclusively for peaceful purposes. The establishment of military bases,
installations and fortifications, the testing of any type of weapons and the
conduct of military manoeuvres on celestial bodies shall be forbidden. The use of
military personnel for scientific research or for any other peaceful purposes shall
not be prohibited. The use of any equipment or facility necessary for peaceful
exploration of the moon and other celestial bodies shall also not be prohibited.''} \\

This clearly rules out unequivocally weapons of any kind located in Earth orbit, and for that matter at any other location in space, and from this, one would infer, the presence of high-power lasers in space. However it stipulates further down that any apparatus designed for peaceful exploration shall NOT be prohibited, and so any space laser, such as a DE-STAR \citep{Lubin2014} structure would seem to be permitted under this article, as the primary intention is entirely benign, i.e. for planetary defence and space exploration.\\ 

For information on the DE-STAR system examined in this paper, go to \cite{Hughes2013,Lubin2014,Kosmo2015,Lubin2015_2,Lubin2016,LUBIN20161093}. In brief, a DE-STAR n laser is a modular phased square array of lasers placed in space where the n indicates the power to which base 10 is raised to give the side of the square array in metres. For example DE-STAR 0 has side $10^0$ = 1 $\si{m}$ (with just a single metre aperture laser) and DE-STAR 1 has side $10^1$ = 10 $\si{m}$ (with 100 lasers of 1 metre aperture) and so on.\\

We find for the DE-STAR 4 option (10 $\si{km}$ $\times$ 10 $\si{km}$), for instance, laser beams of up to 70 $\si{GW}$ can be generated \citep{Lubin2015_2}, thus the possible abuse of such an array for hostile purposes, contrary to article IX of the Outer Space Treaty, needs to be addressed.\\
 
 So are there any strategies which would mitigate or completely nullify the risk of a DE-STAR laser being used malevolently; for example through ground stations being hijacked, or possibly states or private concerns turning rogue? To this end, there are two solutions which stand out, both associated with the laser's location relative to Earth: 
 \begin{enumerate}
      \item{Permanently locate the laser sufficiently far away from the Earth so that the beam flux at Earth is too small to be of any danger.}
      \item{Locate the laser somewhere that is permanently obscured by some intervening celestial body to ensure no line-of-sight.}
 \end{enumerate}
 This paper addresses these two options in specific regard to the DE-STAR phased laser array design.

\section{Specification of the DE-STAR Laser}
\label{sec2}

Table \ref{table:DS_Compare} provides the specifications for the DE-STAR laser arrays considered in this research. Note that the clip ratio is the ratio of minimum waist of each individual laser beam, $w_0$ (defined as the smallest laser spot width for a \textit{Gaussian} beam), and the size of its aperture. We adopt an optimal value of 0.9, as also chosen by \cite{Hettel21}. Note that the power generated column is calculated by scaling up and down from the 70 $\si{GW}$ quoted for the DE-STAR 4 array in \cite{Lubin2015_2, Lubin2016}

\begin{table*}[!h]

\centering
\begin{tabular}{|c|c|c|c|c|}
\hline
DE-STAR & Diameter (m) & Power at 1 au (W) & Clip ratio for each laser & Laser apertures (m) \\ \hline
0 & 1      & 700            & 0.9 & 1 \\
1 & 10     & 70,000         & 0.9 & 1 \\
2 & 100    & 7,000,000      & 0.9 & 1 \\
3 & 1,000  & 700,000,000    & 0.9 & 1 \\
4 & 10,000 & 70,000,000,000 & 0.9 & 1 \\
n & $10^n$    & $70 \times 10^{2n+1}$    & 0.9 & 1 \\ \hline

\end{tabular}

\caption{Specifications adopted for the DE-STAR Phased Array Systems Assumed in this Work} \label{table:DS_Compare}
\end{table*}

\section{Results}
\label{sec3}
\subsection{Minimum Distance from Earth for DE-STAR 0}

\begin{figure}[!htb]
\centering
\includegraphics[scale=0.50]{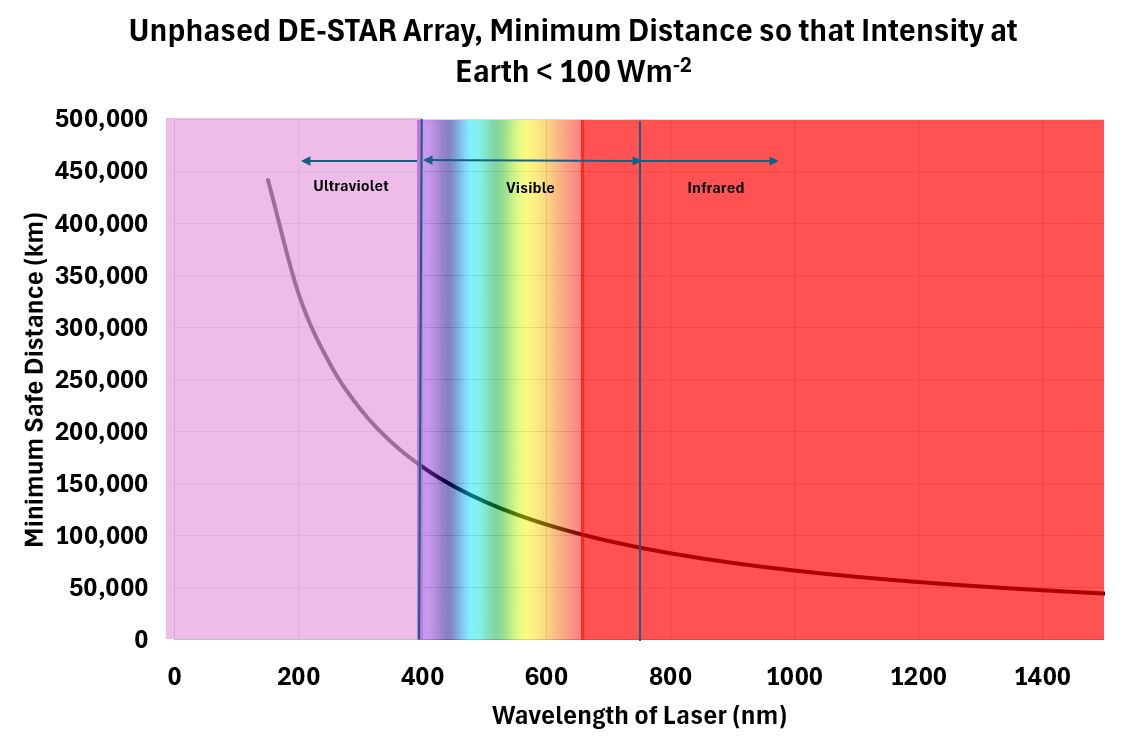}
\caption{Dependence of Minimum Safe Distance of any Unphased DE-STAR Array with Wavelength of the Laser.}
\label{fig:Sing_Aper}
\end{figure}

It is first instructive to determine the minimum level of flux generated by an individual 1 $\si{m}$ aperture laser element (the DE-STAR 0 in Table \ref{table:DS_Compare}). This constitutes the unphased flux of \emph{any} DE-STAR n laser system, because it is independent of 'n'. This can be verified mathematically since the power scales with 'n' as $10^{2n+1}$ and area scales as $10^{2n}$ and so the intensity (defined as power over area) remains fixed.\\

For the theoretical minimum safe distances, we have adopted here a maximum allowed flux (or intensity) at Earth, of $I_{max} = 100$ $\si{Wm^{-2}}$, or on the order 10 $\%$ of the Solar Constant at Earth (1 $\si{au}$ from the Sun).\\

From Table \ref{table:DS_Compare}, we see that the intensity of any unphased DE-STAR beam is 700 $\si{Wm^{-2}}$ (equivalent to DE-STAR 0), i.e. above the previously stated maximum safe intensity threshold. In order that a minimum safe distance from the Earth can be calculated, we now determine how the intensity of an unphased \emph{Gaussian} laser beam reduces with distance.\\

We provide an expression for the intensity of the Gaussian beam, $I_0$, at the point where the spot size $w_0$ is a minimum (the waist of the beam, \cite{Paschottagaussian_beams}):

\begin{equation}
    I_0 = \frac{2P}{\pi w_0^2} 
\end{equation}

The spot size of the laser, $w_z$, varies according to distance, $z$, from the waist, $w_0$, as follows \citep{Paschottagaussian_beams}:
\begin{equation}
    w_z = w_0 \sqrt{1+ \left(\frac{z}{z_r}\right)^2}
\end{equation}
where $z_r$ is the \emph{Rayleigh Length}, given by:
\begin{equation}
    z_r = \frac{\pi n w_0^2}{\lambda}
\end{equation}
, $\lambda$ is the wavelength of the laser beam and $n$, the refractive index is 1.0, for a vacuum.
Furthermore the intensity of the beam is dependent on spot size, $w$, and is:
\begin{equation}
    I_z = \frac{2P}{\pi w_z^2} = I_0\left(\frac{w_0}{w_z}\right)^2
\end{equation}

Combining these equations, and inserting $I_{max}=I_z$, we can derive the minimum safe distance, $z_{min}$ as follows:
\begin{equation}
    \left(\frac{I_0}{I_{max}}\right)^2 = 1+ \left(\frac{z_{min}}{z_r}\right)^2
\end{equation}
\begin{equation}
\label{zminunphased}
    z_{min}=z_r\sqrt{ \left(\frac{I_0}{I_{max}}\right)^2 - 1} = \frac{\pi n w_0^2}{\lambda}\sqrt{ \left(\frac{I_0}{I_{max}}\right)^2 - 1}
\end{equation}

Having derived this relationship \ref{zminunphased}, refer to Figure \ref{fig:Sing_Aper} for a plot of $z_{min}$ over wavelength, and adopting the maximum permissible beam intensity at Earth $I_{max} = 100$ $\si{Wm^{-2}}$.\\

From Figure \ref{fig:Sing_Aper}, we see that as the wavelength of the laser beam increases, so the minimum safe distance reduces. For instance, an infrared laser would need to be just outside geosynchronous Earth orbit (GEO) ($\sim{35,000} \si{km}$). On the other hand, at the other extreme an ultraviolet laser with $\lambda \sim{200} \si{nm}$, would have to be be outside cis-lunar space. Note here that the results provided in Figure \ref{fig:Sing_Aper} are, by definition, the extreme minimum distances, and when phasing is considered the minimum safe distances increase, read on.\\

\begin{figure*}[!h]
\centering
\hspace{-1.0cm}
\includegraphics[scale=0.50]{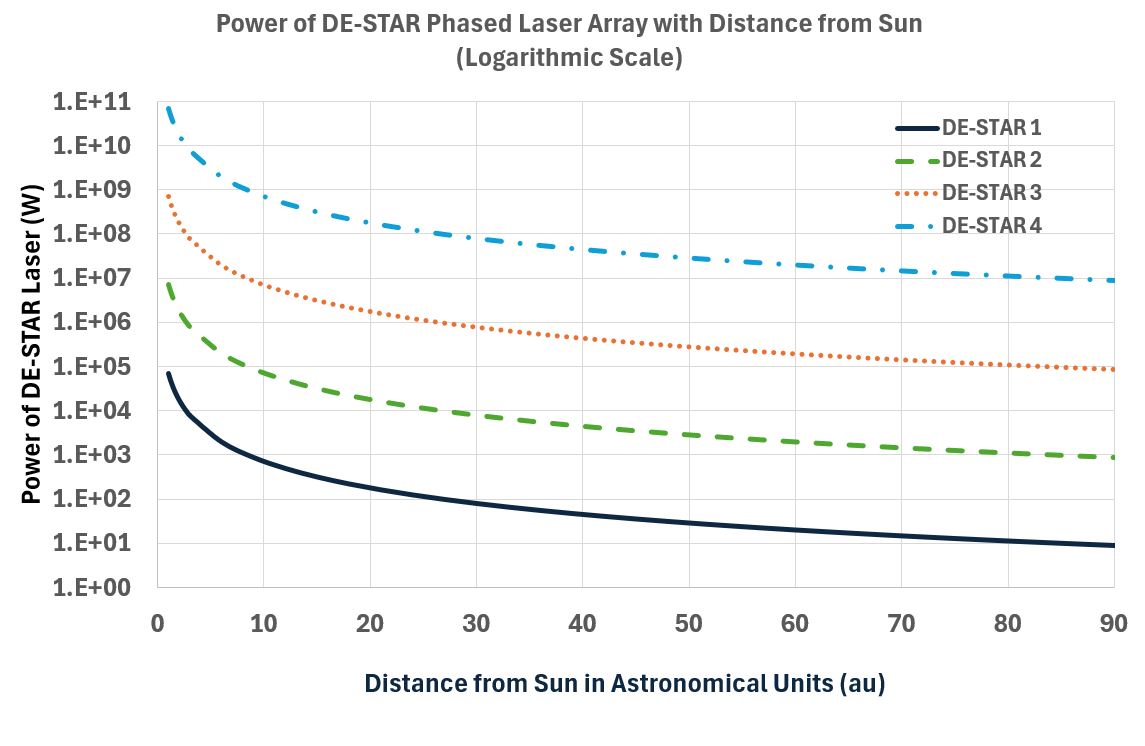}
\caption{Decay in Power of DE-STAR 1 to DE-STAR 4 According to Distance from the Sun (Logarithmic Scale).}
\label{fig:Pow_Dec}
\end{figure*}

\subsection{Reduction in Power Generated by DE-STAR with Sun Distance}

 We shall shortly address the case where the DE-STAR laser array is operated as a phased array, where considerably higher intensity levels can be expected. However before so proceeding, we first examine the reduction in power of a DE-STAR laser array as it is positioned farther away from the Sun. This reduction is a consequence of the decrease in solar flux intensity on the photovoltaic cells, where an inverse square law is followed. The plot is provided in Figure \ref{fig:Pow_Dec}. Note that to generate this Figure, we yet again scale the array power according to a DE-STAR 4, assuming that the 70 $\si{GW}$ it generates is at 1 $\si{au}$ from the Sun.\\

We find that a DE-STAR n at 90 $\si{au}$ from the Sun is approximately equivalent to a DE-STAR n-1 at 10 $\si{au}$ and a DE-STAR n-2 at 1 $\si{au}$. \\

\subsection{Minimum Distance from Earth for DE-STAR n}
We now can determine the minimum safe distance of a DE-STAR when it is behaving as a phased array.\\

For any DE-STAR like array of diameter D and wavelength $\lambda$, the central lobe for the beam remains contained within an approximate angle $\theta$, given as follows \citep{Hettel21}:
\begin{equation}
    sin\ \theta < \frac{D}{\lambda} \implies \theta \lesssim  \frac{D}{\lambda}
\end{equation}
We shall now assume that all of the power is confined within this spread of angles \citep{Lubin2016}. The area encompassed by this angle at a distance $z$ from the DE-STAR is given as follows:
\begin{equation}
    Area = \pi (z\theta )^2 = \pi \left(\frac{zD}{\lambda}\right)^2
\end{equation}

Thus we find the intensity of the laser a distance $z$ away is as follows:
\begin{equation}
    I_z = \frac{P}{Area}= \frac{P}{\pi}\left(\frac{\lambda}{zD}\right)^2
\end{equation}

We now rearrange and make $I_z = I_{max}$:

\begin{equation}
\label{zmin}
    z_{min}=  \frac{D}{\lambda} \sqrt{\frac{P}{\pi I_{max}}}
\end{equation}

We find when we apply the above equation to DE-STAR 3 and DE-STAR 4 laser arrays, the values of $z_{min}$ become $\gtrsim{2}$ $\si{au}$. We observe here that any phased laser array positioned such distances from Earth will also be significantly displaced from the Sun, reducing the photovoltaic output as indicated in Figure \ref{fig:Pow_Dec}. The following derives an approximate correction to equation \ref{zmin}.\\

First we scale the power generated by distance $z$ as follows:
\begin{equation}
    P_z = P \left(\frac{1 au}{z}\right)^2
\end{equation}

Inserting this into equation \ref{zmin} we get:

\begin{equation}
\label{zmin2}
z_{min}=  \frac{D}{\lambda} \sqrt{\frac{P_z}{\pi I_{max}}} = \frac{D\ au}{\lambda z_{min}}\sqrt{\frac{P}{\pi I_{max}}} \implies    z_{min} = \sqrt{\frac{D\ au}{\lambda}\sqrt{\frac{P}{\pi I_{max}}}} 
\end{equation}

\begin{figure}[!htb]
\centering
\includegraphics[scale=0.5]{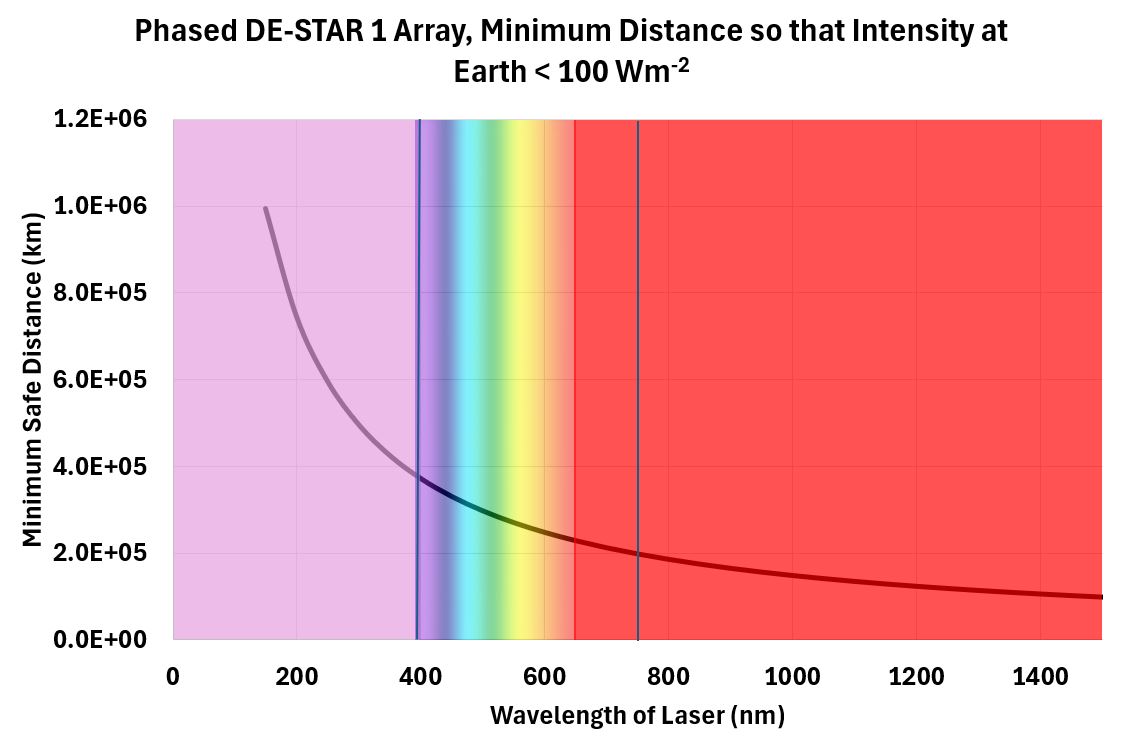}
\caption{DE-STAR 1}
\label{fig:DS1}
\end{figure}

\begin{figure}[!htb]
\centering
\includegraphics[scale=0.5]{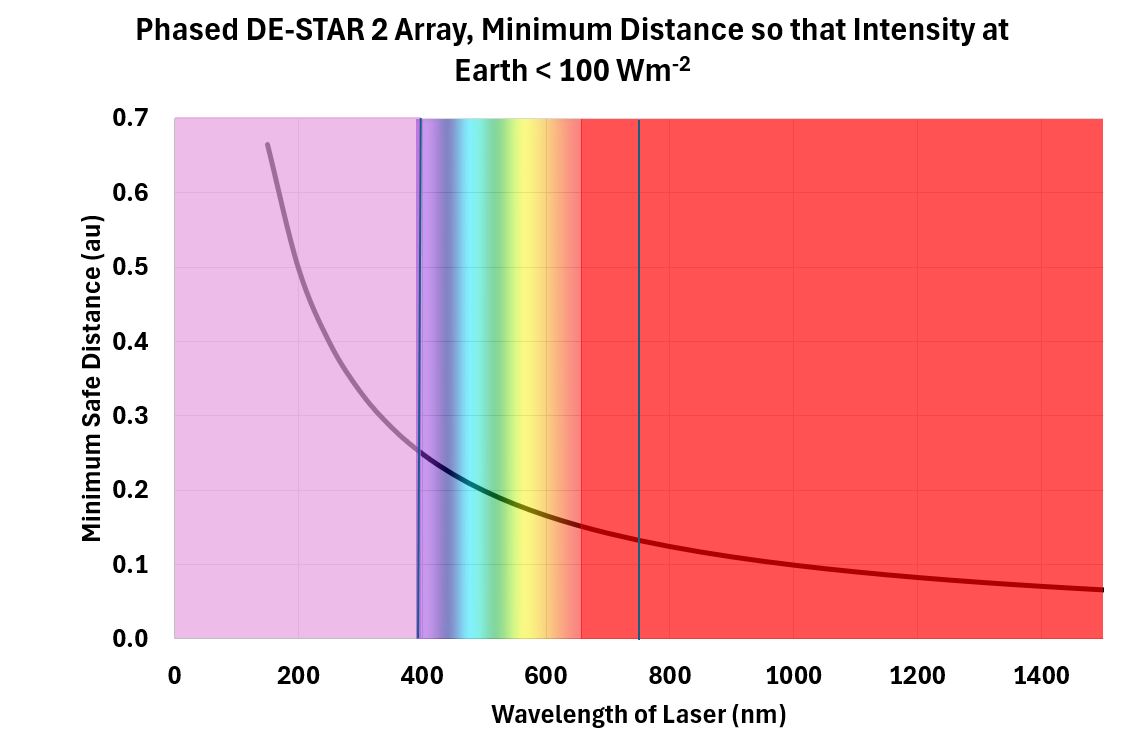}
\caption{DE-STAR 2}
\label{fig:DS2}
\end{figure}

\begin{figure}[!htb]
\centering
\includegraphics[scale=0.5]{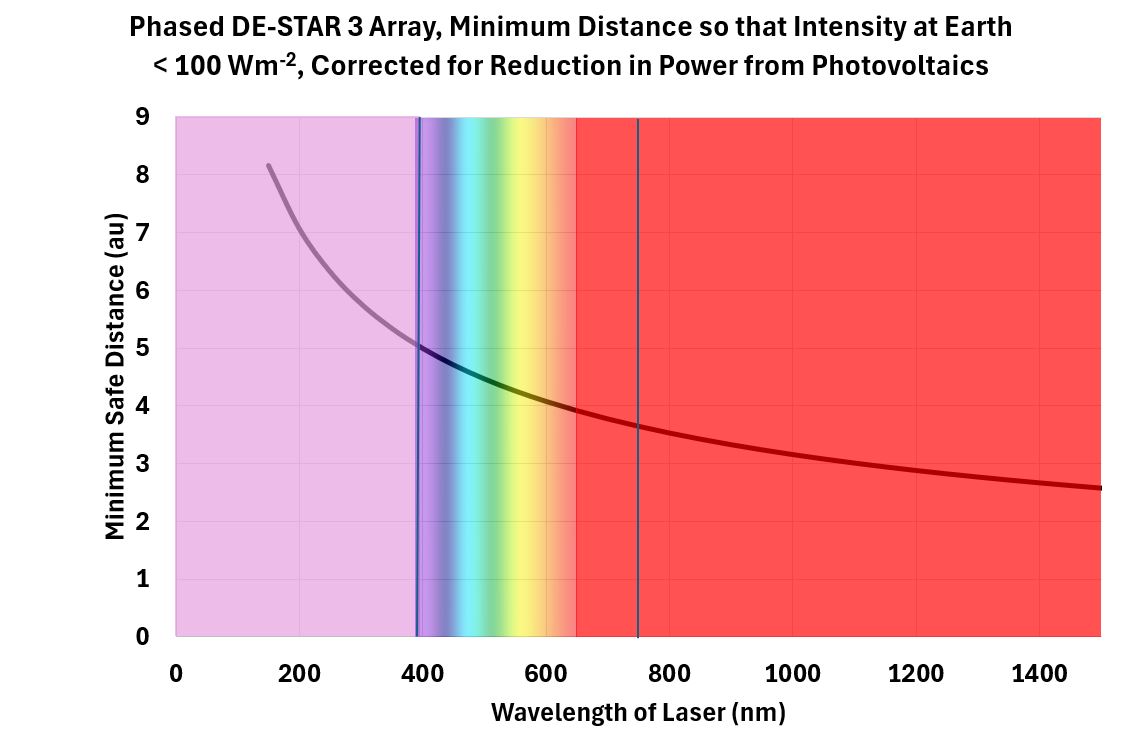}
\caption{DE-STAR 3 Corrected for Reduced Photovoltaic Output}
\label{fig:DS3_C}
\end{figure}

\begin{figure}[!htb]
\centering
\includegraphics[scale=0.5]{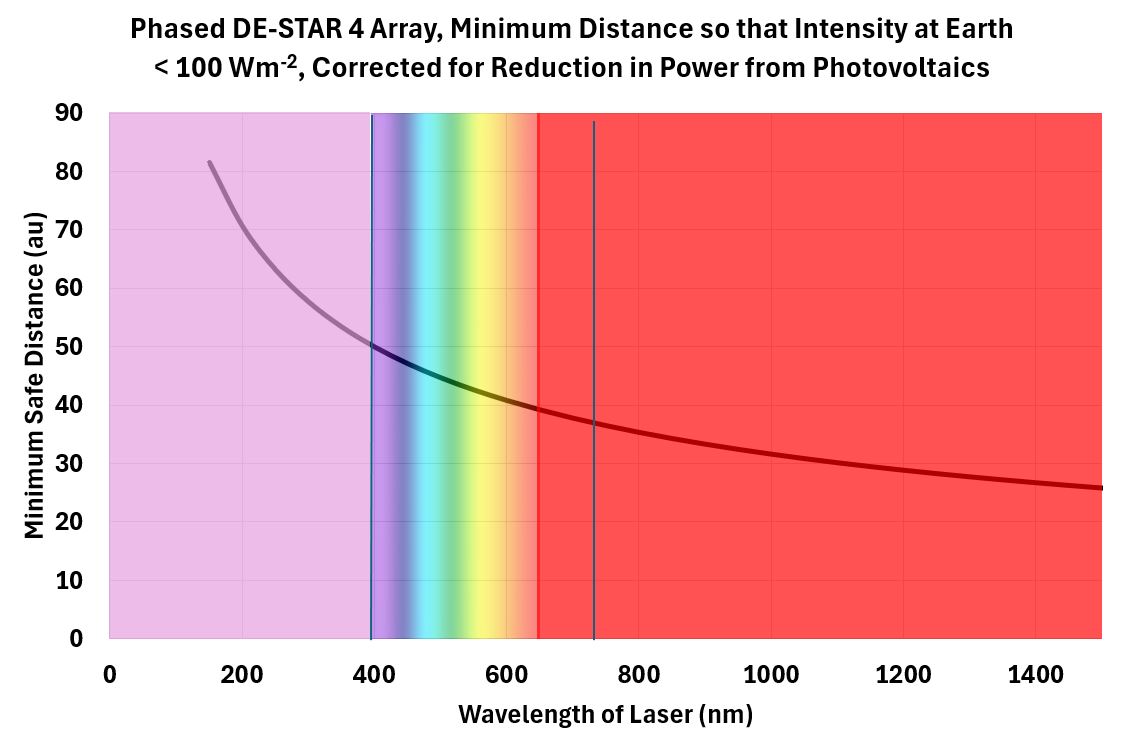}
\caption{DE-STAR 4 Corrected for Reduced Photovoltaic Output}
\label{fig:DS4_C}
\end{figure}

The $z_{min}$ plots of equation \ref{zmin} for DE-STAR 1 \& 2 and the corrected equation \ref{zmin2} for DE-STAR 3 \& 4 are provided in Figures \ref{fig:DS1} \& \ref{fig:DS2} and Figures \ref{fig:DS3_C} \& \ref{fig:DS4_C} respectively.

\label{app1}

\section{Discussion}
\label{sec4}
From Figure \ref{fig:DS1} we find for a simple DE-STAR 1, 10m x 10m square array with wavelength, $\lambda$, for most $\lambda$ it would have to be placed outside of cis-lunar space to remain safe to operate , and for ultraviolet $\lambda  \lesssim 350 \si{nm}$, it would have to be even further away - beyond the Earth/Moon Lagrange 2 point, which is around 450,000 $\si{km}$ from Earth. A Sun-Earth L2 location, however, would be fine for any DE-STAR 1 array.\\

From Figure \ref{fig:DS2}, for a DE-STAR 2, we find safe distances are around $z_{min} \sim 0.1\si{au} - 0.7 \si{au}$. Thus, for a DE-STAR 2, the laser could, for example, orbit 1 $\si{au}$ from the Sun at an angular displacement of $\sim{41}$ $\si{deg}$ from Earth, either in advance or behind Earth in its heliocentric orbit. As these are unstable points in the Sun/Earth system, a preferable location would be 60 $\si{deg}$ either side of Earth, which are the stable Sun/Earth Lagrange 4 and 5 points where the laser will maintain a fixed location w.r.t. Earth with no on-board propulsion necessary.\\

 Referring to Figure \ref{fig:DS3_C}, we find that for a DE-STAR 3, taking into account the reduction in power supplied by the photovoltaics at larger distances from the Sun, it would have to be placed somewhere beyond the asteroid belt (in fact $\gtrsim{3} \si{au}$). Further if the chosen $\lambda$ is in the ultraviolet, the DE-STAR 3 would have to be positioned beyond the orbit of Jupiter ($\gtrsim{5} \si{au}$). Note that this considerably reduces the power output of the photovoltaic cells. \\
 
Refer now to Figure \ref{fig:Pow_Dec}, which reveals on a logarithmic scale how the power of a DE-STAR 1-4 would drop according to Sun-distance. As already stated, we find that if a DE-STAR n is placed a distance of 90 \si{au} from the Sun, then that is equivalent to a DE-STAR n-1 at 10 \si{au} or a DE-STAR n-2 at 1 \si{au} from the Sun. From this we can conclude that, ignoring the logistical difficulties in locating a DE-STAR 3 there, nonetheless there is a considerable performance advantage to placing a DE-STAR 3 in a Jupiter orbit, as compared to a DE-STAR 2 at 1 $\si{au}$ (refer Figure \ref{fig:Pow_Dec}).\\

Finally the DE-STAR 4 phased array in Figure \ref{fig:DS4_C} would have to be placed at least $\sim{30 - 40} \si{au}$, and possibly above $\sim{70}$ $\si{au}$ (for ultraviolet) from the Sun to remain safe, but nevertheless, as we have already remarked (refer Figure \ref{fig:Pow_Dec}), this would still generate more power than a DE-STAR 2 at 1 $\si{au}$, but a DE-STAR 3 at the orbit of Jupiter ($\sim{5}$ $\si{au}$) would generate more power.\\

Note that all the above analysis assumes there is a direct line-of-sight between the DE-STAR array and Earth, however, as mentioned in Section \ref{sec1}, there are certain locations in the Solar System where ANY such array could reside with absolutely NO line-of-sight and what’s more there would be no dependence on the ‘n’ in DE-STAR n. \\

Two such locations are the Earth/Moon Lagrange 2 point  (on a line from the Earth to the Moon, extending beyond the Moon by $\sim{61,000}$ $\si{km}$) and the Sun/Earth Lagrange 3 point (at 1 $\si{au}$ from the Sun and diametrically opposite the Earth as it orbits the Sun). In both cases, the instability of these points will result in the DE-STAR wandering away and potentially becoming visible from Earth, so an on-board propulsion would be needed to prevent this. One solution would be to use the push-back from the lasers to provide a means of corrective propulsion. However it would appear a DE-STAR's placement at either of these points is not an entirely satisfactory solution to the problem.\\

Note also that should the DE-STAR 3 or 4 need to be used to their full potential, and safety considerations were deemed important, then strategies other than those addressed above could be adopted. One such solution might be to physically constrain the orientation of the laser so that it can never be pointed at Earth. The means by which this can be achieved, either via hardware or unmodifiable embedded software, or other, is beyond the scope of this paper.\\

\section{Conclusion}
\label{sec5}
We considered the potential dangers of situating in space, a powerful laser of the DE-STAR specification conceived by Lubin etc. We found that although there are locations in the Solar System (Sun/Earth L3 and Earth/Moon L2) where there is no direct line-of-sight to Earth, these points tend to be unstable and are therefore NOT a permanent solution to the problem.\\

However we found placing the DE-STAR at considerable distances from the Earth can ensure that the beam intensity drops below a fiducial maximum safe flux of 100 $\si{Wm^{-2}}$ at Earth.\\

Thus for a DE-STAR 2 placed at 1 $\si{au}$ from the Sun, it could orbit in advance or behind the Earth by $\sim{41}$ $\si{deg}$ and it would be safe, but ideally it would instead be placed at one of the Sun/Earth Lagrange 4 or 5 points, as these are stable orbits.\\

As for all lasers considered, the minimum safe distance for a DE-STAR 3 depends upon the exact wavelength of the laser beam, but would have to be at least beyond the Asteroid Belt (infrared), and even further out to beyond Jupiter's orbit in the ultraviolet, i.e. for $\lambda <$ 400 $\si{nm}$.\\

DE-STAR 4 \& 5 are extremely powerful lasers, and in principle would have to be located considerable distances from the Sun to insure no danger to Earth, or alternatively other solutions would have to be considered.




\bibliography{Safe_DE-STAR}{}
\bibliographystyle{aasjournal}
\end{document}